\documentclass[twocolumn]{aastex631}

\usepackage{xspace}
\usepackage{amsmath}

\newcommand{\Rstar}{\ensuremath{R_{\star}}\xspace}

\newcommand{\Rearth}{\ensuremath{R_\oplus}\xspace}

\newcommand{\Teff}{\ensuremath{T_\mathrm{eff}}\xspace}
\newcommand{\logg}{\ensuremath{\log{g}}\xspace}
\newcommand{\feh}{\ensuremath{\mathrm{[Fe/H]}}\xspace}
\newcommand{\afe}{\ensuremath{\mathrm{[}\alpha\mathrm{/Fe]}}\xspace}
\newcommand{\mh}{\ensuremath{\mathrm{[M/H]}}\xspace}

\newcommand{\Gaia}{\textit{Gaia}\xspace}
\newcommand{\Kepler}{\textit{Kepler}\xspace}
\newcommand{\gspspec}{\texttt{GSP-Spec}\xspace}
\newcommand{\aroche}{\ensuremath{a_\mathrm{R}}}
\mathchardef\mhyphen="2D
\begin{document}
\title{The Period Distribution of Hot Jupiters is Not Dependent on Host Star Metallicity}

\author[0000-0001-7961-3907]{Samuel W.\ Yee}
\author[0000-0002-4265-047X]{Joshua N.\ Winn}
\affiliation{Department of Astrophysical Sciences, Princeton University, 4 Ivy Lane, Princeton, NJ 08544, USA}

\begin{abstract}
The probability that a Sun-like star has a close-orbiting giant planet
(period~$\lesssim$~1~year) increases
with stellar metallicity.
Previous work provided evidence that the period distribution
of close-orbiting giant planets is also linked to metallicity,
hinting that there two formation/evolution pathways
for such objects, one of which is more probable
in high-metallicity environments.
Here, we check for differences
in the period distribution of hot Jupiters ($P < 10$~days) 
as a function of host star metallicity,
drawing on a sample of 232 transiting hot Jupiters
and homogeneously-derived metallicities from 
\textit{Gaia} Data Release 3.
We found no evidence for any metallicity dependence; the period distributions of hot Jupiters around metal-poor and metal-rich stars are indistinguishable.
As a byproduct of this study, we provide transformations
between metallicities from the \textit{Gaia} Radial Velocity Spectrograph
and from traditional high-resolution optical spectroscopy
of main-sequence FGK stars.
\end{abstract}

\section{Introduction} \label{sec:intro}

One of the goals of exoplanet demographics -- measuring planet
occurrence rates and the statistical distributions of their properties -- is to improve our understanding of planet formation. One of the earliest
demographic discoveries was the strong
and positive association between stellar metallicity and giant-planet
occurrence \citep[][]{Gonzalez1997,Santos2004,Fischer2005}.
This ``metallicity effect'' is often interpreted as evidence in favor of the core-accretion theory of planet formation,
and is observed to be stronger for hot Jupiters than for smaller or longer-period
planets \citep{Petigura2018}.

Another important topic
in giant-planet demographics is
the measurement of their
orbital period distribution.
Early results from the radial-velocity surveys \citep[e.g.,][]{Udry2003,Butler2006,Cumming2008}
found that the period distribution
has a peak at 3--5 days, falling at intermediate periods
followed by a gradual
rise as the period increases further.
Early transit surveys found that the very shortest-period giant planets ($P < 3$~days)
are rarer than those with periods between
3 and 10 days \citep[e.g.,][]{Gaudi2005}, after correcting for detection biases.
These results gave rise to the notion of a ``three-day pile-up'' in the period distribution
of giant planets,
Later,
\citet{Santerne2016} used the {\it Kepler}
data to detect a 2-$\sigma$
peak centered on $P\approx$~3--5 days
in the giant-planet occurrence rate, supporting
the general notion of a ``pile-up'',
although we caution that this term has
not been used consistently in the literature.
Some authors have in mind a narrow peak
at 3-days wholly contained within the $P<10$~day
range of hot Jupiters, while others
are referring to a general excess
of hot Jupiters relative to
warm Jupiters.

\cite{Dawson2013} had the insight to search for a connection
between the metallicity effect and the giant-planet period distribution.
They examined all of the giant-planet candidates
with $P \lesssim 500$~days that had been detected by NASA's \Kepler mission,
and examined the period distributions
separately for stars with super-solar
and sub-solar metallicities. They found that the distributions differed significantly,
reporting that ``only metal-rich stars host a pile-up of hot Jupiters.''
These findings,
along with evidence that high eccentricities
of warm Jupiters ($0.1 < a < 1$~AU)
are associated with high metallicities, led
\cite{Dawson2013} to propose that both disk-driven
migration and high-eccentricity migration produce hot Jupiters.
In this picture, disk-driven migration does not produce a period
pile-up and depends weakly if at all on metallicity;
high-eccentricity migration produces 
a pile-up of giant planets at short orbital periods
and occurs more often in metal-rich disks, where multiple
giant planets are likelier to form and trigger a dynamical instability.

Although the statistical
tests performed by \cite{Dawson2013} were
based on giant planets with periods
ranging out to 500 days,
if their interpretation is correct,
one might expect differences between the metal-rich and metal-poor samples
even when restricting the period range to
$P < 10$~days.
With this in mind,
\citet{Hellier2017} examined
period distribution of all the transiting giant planets with $P < 22$~days that had been
confirmed with mass measurements and with measured host star metallicities.
Their sample of 271 planets consisted mainly of hot
Jupiters discovered by the ground-based transit surveys, as opposed to {\it Kepler},
because most of the {\it Kepler} hot Jupiters
had not been confirmed (and remain unconfirmed
today).
They found no evidence for any statistical difference between the period distributions
of giant planets orbiting metal-rich and metal-poor stars within this period range.
However, the sample was drawn from
a heterogeneous collection of surveys, involved
a wide range of stellar types,
and adopted metallicities derived by
many different authors.

We and others
have been assembling a magnitude-limited sample of several hundred transiting hot Jupiters
($P < 10$~days), by combining the past two decades of discoveries with new discoveries from the NASA Transiting Exoplanet
Survey Satellite \citep[TESS;][]{TESS_Ricker15}.
Our contribution
goes by the name of the Grand Unified Hot Jupiter Survey
\citep{Yee2022a,Yee2022b}.
Objects in this sample orbit brighter stars and have been
vetted more thoroughly than the Kepler sample, while
being more complete and homogeneous than the collection studied
by \citet{Hellier2017}.
Spectroscopic metallicities are available
in almost all cases from ground-based high-resolution optical spectroscopy
and/or from the Radial Velocity Spectrometer (RVS)
aboard the ESA Gaia spacecraft \citep{GaiaMission_2016,GaiaRVS_Cropper2018,GaiaDR3Summary_GaiaCollaboration2022a}.
Given the increased size, reliability and completeness (\S\ref{ssec:completeness}) of the currently available
sample,
we decided to 
revisit the relationship
between the metallicity effect and the period distribution of hot Jupiters.

\begin{figure*}
\plottwo{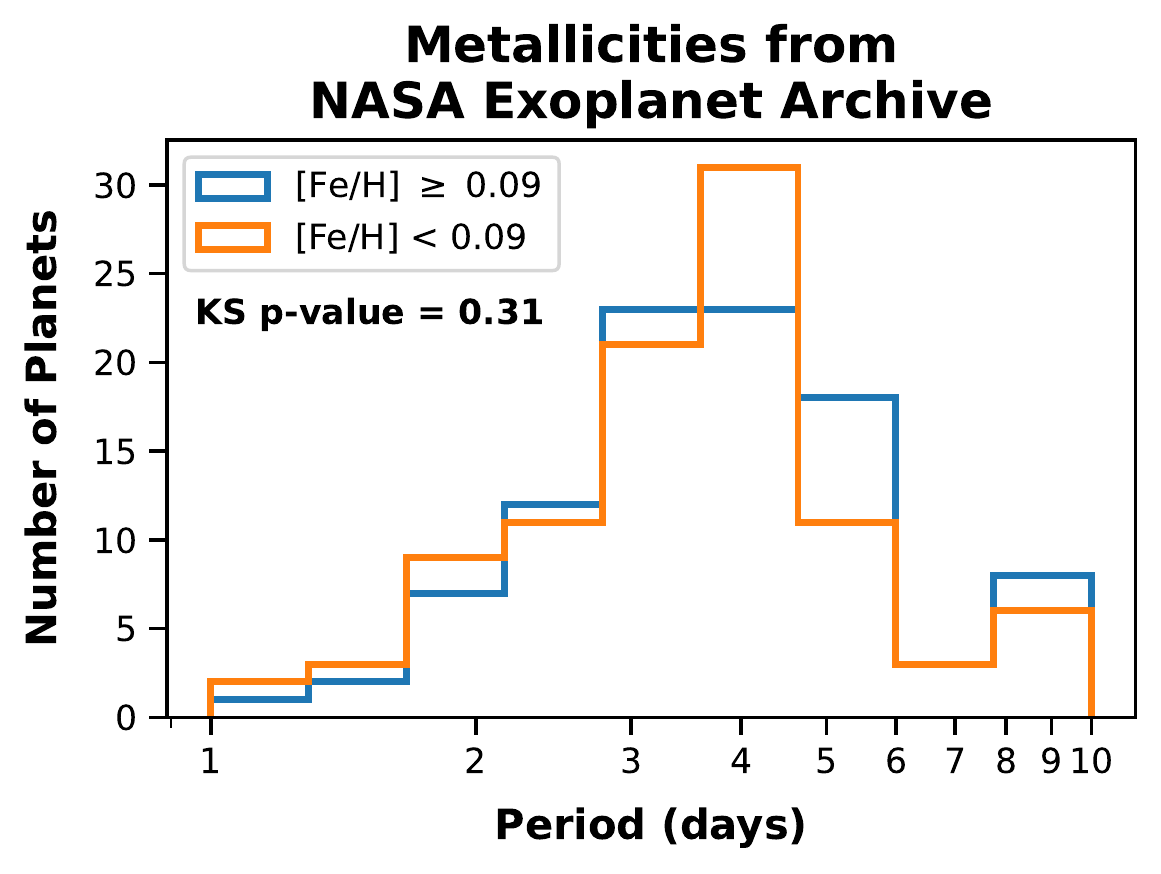}{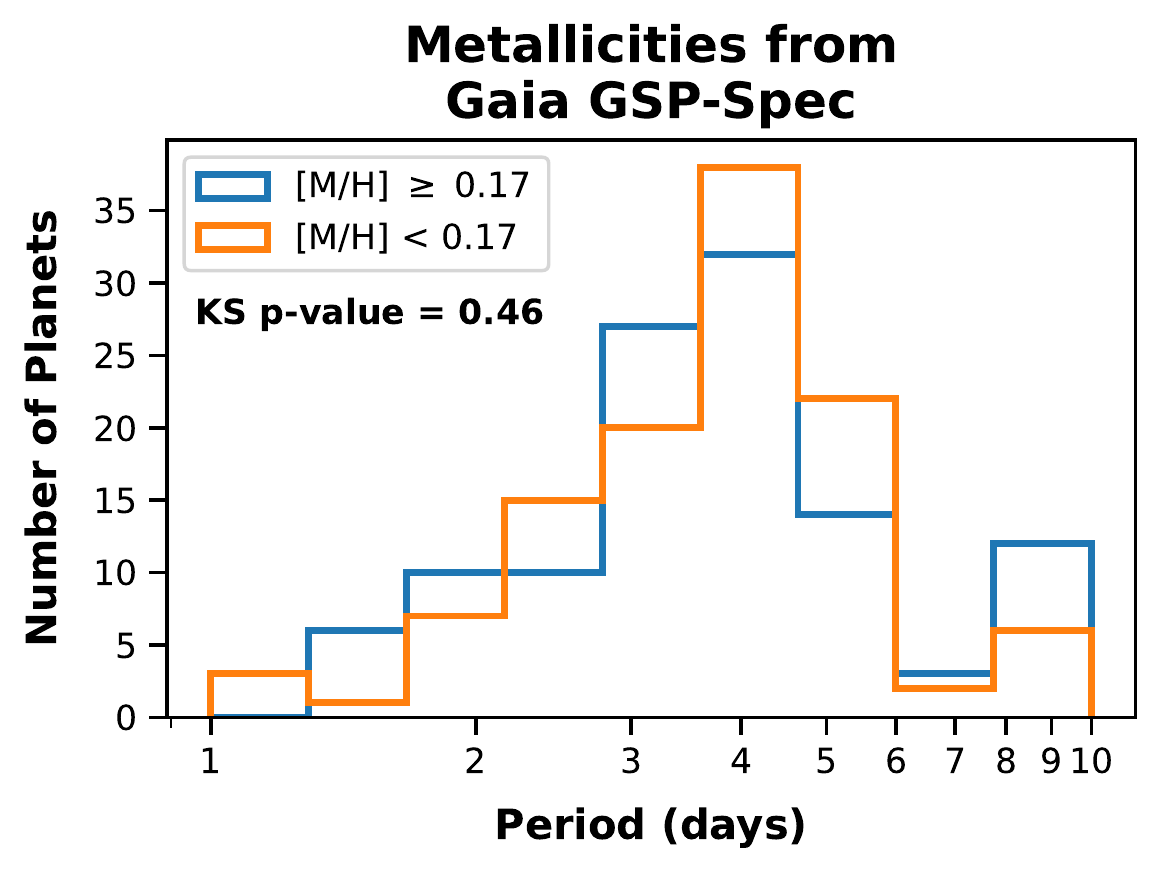}
\caption{\textbf{Left:} Period distribution of confirmed hot Jupiters in the
NASA Exoplanet Archive orbiting FGK stars brighter than $G = 12.0$~mag, split by metallicity relative to the median.
The metallicities were drawn from
the literature (\S\ref{sec:exoarchive}).
\textbf{Right:} Period distribution of an enlarged sample of hot Jupiters with
$G<12$, split by metallicity relative to the median.
In this case, the metallicities are
based on calibrated \Gaia spectroscopic (GSP-Spec)
metallicities (\S\ref{sec:homogeneous_sample}).
In both cases, the observed distribution
has a peak at 3--5 days, although we caution that this does not necessarily
correspond to a peak in the intrinsic occurrence rate; the distributions were not corrected for detection efficiency.
K-S tests do not
reject
the hypothesis
that the period distributions of the metal-poor
and metal-rich samples are drawn from the
same distribution.
\label{fig:period_dist}}
\end{figure*}

\section{A Large Heterogeneous Sample} \label{sec:exoarchive}

For a first glimpse at the results, 
we used data from the \cite{ExoplanetArchive_PSCompPars} (NEA).\footnote{\url{https://exoplanetarchive.ipac.caltech.edu/}}
We queried the database
on November 2, 2022 for
transiting hot Jupiters, defined as a planet with a radius
between 8 and 24\,$\Rearth$ and an orbital period $P < 10$~days.
We also required the star
to have an effective temperature between 4400 and 6600\,K, a radius smaller than 2.2$\,R_\odot$,
and an apparent Gaia magnitude brighter than 12, the limiting
magnitude for which completeness is expected to be high \citep{Yee2021b}.
This resulted in a sample of 202 confirmed transiting hot Jupiters
around main-sequence FGK stars.

Using the ``default''
NEA metallicities, which are drawn from the various sources in the literature,
we split the sample into low- and high-metallicity
subsamples relative to the median metallicity of 
[Fe/H]$=+0.09$~dex. The left panel of Figure \ref{fig:period_dist} shows the orbital period distribution of
each subsample.
Visually, the two distributions do not seem significantly different.
Both distributions exhibit a peak at 3--5 days,
although it is premature to conclude that the intrinsic period
distribution has any such peak, because the transit detection
efficiency declines steeply with increasing period \citep[see, e.g.][]{Gaudi2005}.
A two-sample Kolmogorov-Smirnov (K-S) test yielded a $p$-value of 0.31, and therefore cannot reject the null hypothesis
that the orbital periods are drawn from the same parent
distribution.
We repeated the same test after
splitting the sample at $\feh = 0.0$~dex,
instead of the median metallicity, to match the
breakpoint chosen by \citet{Dawson2013} and \citet{Hellier2017}.
Again, we found that the period distributions are indistinguishable
via the K-S test, with $p=0.90$.\\[0.1in]

\section{A More Homogeneous Sample} \label{sec:homogeneous_sample}

Next, we assembled a more homogeneous dataset of host star metallicities (\S\ref{ssec:gaia_metallicities}), and considered the effects of sample incompleteness (\S\ref{ssec:completeness}).

\subsection{Gaia Metallicities} \label{ssec:gaia_metallicities}

\begin{figure*}
\plotone{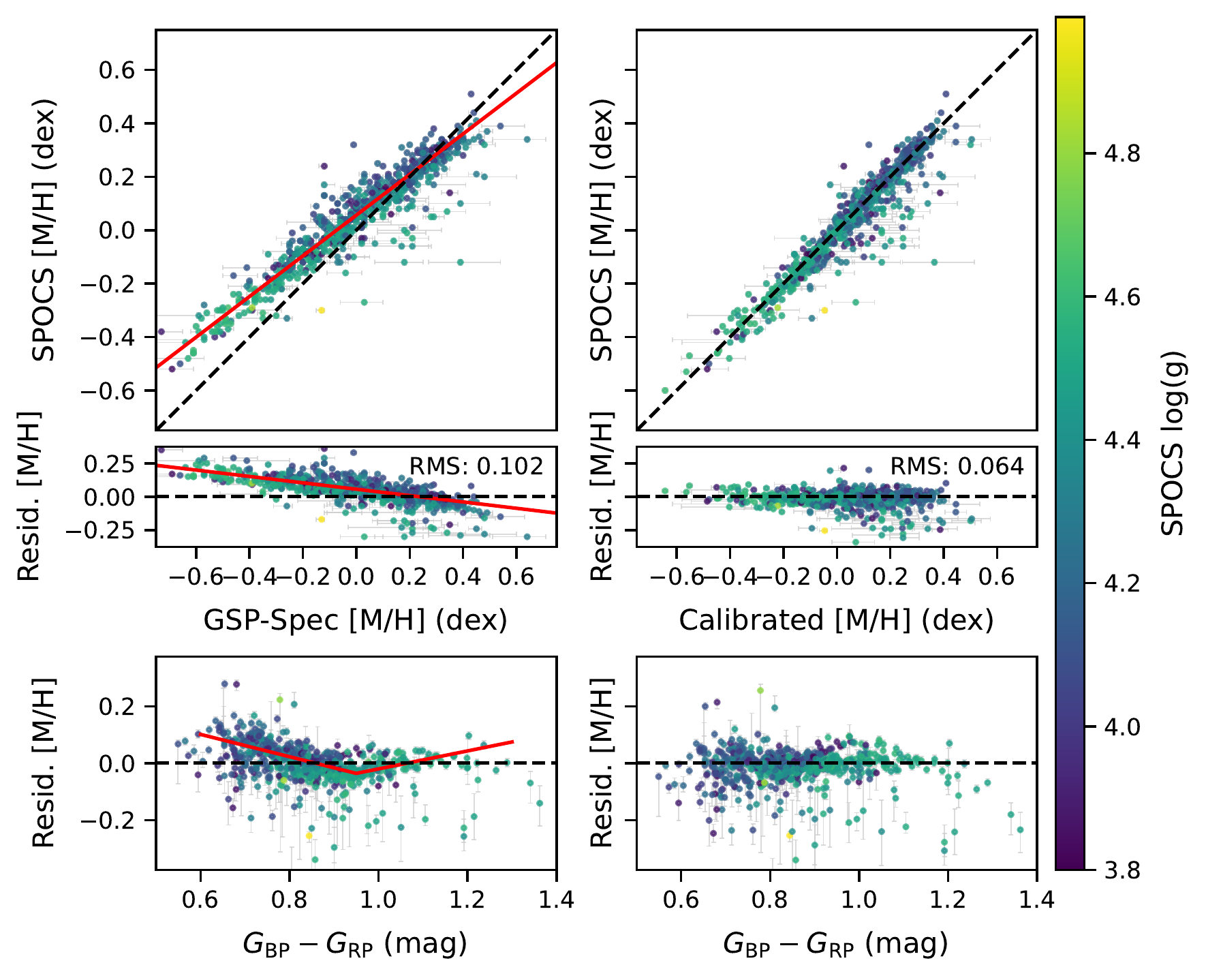}
\caption{The \Gaia\ \gspspec metallicities are biased as a function of the
inferred metallicity (top left panel). After fitting a linear function to the
residuals between the SPOCS and \Gaia \mh values, the remaining residuals still
exhibit a trend as a function of \Gaia  $G_\mathrm{BP} - G_\mathrm{RP}$ color
(bottom left panel). We removed this trend by fitting a piecewise linear
function with a breakpoint at 0.95. Using the calibration coefficients from
a joint fit to these variables, the rms residual between the SPOCS and calibrated
\Gaia \mh values is 0.064 dex (right panels).
\label{fig:gaia_mh_calibration}}
\end{figure*}

The NEA metallicities were derived using many
different instruments and analysis methods. 
Given the varying resolution, spectral coverage, and signal-to-noise
ratios of the spectroscopic observations, and the variety of codes 
and spectral libraries that were employed,
we expect systematic offsets on the order of 0.1--0.2~dex
that prevent more precise comparisons between
metallicities in the sample \citep[e.g.,][]{CKSI_Petigura2017}.
Some metallicities were also derived
from the less accurate method of fitting
stellar-evolutionary models to broad-band photometric measurements.


\Gaia Data Release 3 (DR3; \citealp{GaiaDR3Summary_GaiaCollaboration2022a})
provides a new and more homogeneous source of spectroscopic metallicities.
The RVS instrument is a spectrograph with $R \approx 11{,}500$
covering the wavelength range 846--870\,nm \citep{GaiaRVS_Cropper2018}.
Stellar atmospheric parameters are available
for about 5.6 million stars,
including almost all of the stars brighter than $G = 12$~mag
\citep{GaiaDR3_GSPSpec_Recio-Blanco2022}.
The \Gaia DR3 \texttt{AtmosphericParameters} table contains \Teff, \logg,
\mh, \afe, and up to 15 individual elemental abundances derived from the
\texttt{GSP-Spec\,MatisseGaugin}
workflow, which used an iterative procedure to match the observed
spectra to a grid of synthetic templates. Confidence intervals for these parameters
were determined through Monte-Carlo resampling of the flux uncertainties on
each wavelength pixel.

The large size and uniformity of this catalog made it
attractive for our work, but first
we checked for
biases and offsets relative to other spectroscopic catalogs. \citet{GaiaDR3_GSPSpec_Recio-Blanco2022} found that the RVS \logg and \mh determinations
have a gravity-dependent bias
and proposed polynomial corrections based on comparisons with the APOGEE,
GALAH, and RAVE spectroscopic surveys. We decided to perform our own
calibration focused on the main-sequence FGK stars.

We took as our ground truth the Spectroscopic Properties
of Cool Stars (SPOCS) sample from \citet{Brewer2016}, who
analyzed high-resolution, high signal-to-noise Keck/HIRES spectra of
1{,}617 FGK stars. They used the same line list in all cases,
and fitted each spectrum
with the same code \citep[SME,][]{SME_Valenti1996} to
derive spectroscopic properties and abundances for these targets.
They achieved
precisions of 25\,K in \Teff, 0.01\,dex in \mh, and 0.028\,dex in \logg, as
determined from separate observations of the same star. For this analysis, we restricted the SPOCS sample to the 1{,}275 stars with $\logg \geq 3.8$.

We 
downloaded the \gspspec atmospheric parameters for these stars from the
Gaia archive.\footnote{\url{https://gea.esac.esa.int/archive/}}
Gaia parameters were available for
1{,}150, or 90\%, of the SPOCS sample.
In some cases,
the parameters had quality flags indicating possible problems.
We imposed the strictest quality cuts, requiring 
all quality flags to be 0 except for the \texttt{fluxNoise}
flag which was allowed to be 0 or 1
(see \citet{GaiaDR3_GSPSpec_Recio-Blanco2022} for details on each flag). 
This left 677 remaining stars.

After some experimentation, we found that the best calibration
was achieved by fitting
the SPOCS \mh values with a combination of a linear function of the \gspspec \mh and a
piecewise linear function of the $G_\mathrm{BP} - G_\mathrm{RP}$ color
(Figure \ref{fig:gaia_mh_calibration}):
\begin{align}
\mh_\mathrm{cal} =\,&a + b \mh_\mathrm{GSP\mhyphen Spec} + c (\mathrm{BP\text{--}RP}) \\
&+ d \left(\mathrm{BP\text{--}RP} - 0.95\right) \cdot H(\mathrm{BP\text{--}RP} - 0.95),\notag
\end{align}
where $H(x)$ is the Heaviside function. Thus, the breakpoint of the piecewise
linear function is at $G_\mathrm{BP} - G_\mathrm{RP} = 0.95$.
The need for a broken linear correction in color is likely due to
temperature-dependent systematic differences in the model atmospheres used, with the
breakpoint corresponding roughly to the color stars of solar temperature.
The best-fit calibration coefficients are $a = 0.384$, $b=0.761$, $c=-0.392$,
and $d = 0.708$.
Applying this calibration reduced
the root-mean-squared (rms) difference between SPOCS and \gspspec metallicities
from 0.10~dex to 0.064~dex. We saw
no correlations in the residuals with the SPOCS \Teff or
\logg parameters. This calibration was used to
correct the \Gaia\ \gspspec parameters for our sample of hot Jupiter hosts,
providing a more homogeneous catalog of spectroscopic
metallicities.

\subsection{Sample Completeness} \label{ssec:completeness}

Most of the hot Jupiters in the NEA
were discovered in wide-field ground-based transit surveys
that were subject to complex selection effects, such as
a strong bias favoring the detection
of short-period planets \citep{Gaudi2005}.
In addition, the parent population of stars that were thoroughly
searched for hot Jupiters is not well characterized.
In \citet{Yee2021b}, we simulated the expected characteristics of a
magnitude-limited sample, based on \Kepler statistics,
and found that the collection of hot
Jupiters known at the time was probably only $\approx$45\% complete
down to $G=12$. A sample with low completeness is still acceptable
for comparing the period distributions of metal-poor
and metal-rich stars are still valid and interpretable, but only if
the selection effects do not depend
on metallicity. 
For example, although the geometric transit probability
distorts the observed period distribution away from the intrinsic
distribution, this applies equally to all planets, and will not
affect our comparison apart from determining the overall number
of planets in the sample.

However, it is important to check on the possibility that
our ability to detect a transiting giant planet at fixed orbital period
and apparent magnitude depends on the host star's metallicity.
For this purpose, we used data from NASA's Transiting
Exoplanet Survey Satellite (TESS; \citealt{TESS_Ricker15}).
We downloaded all the light-curves generated by the
TESS Quick Look Pipeline (QLP; \citealt{TESS_QLP_Huang2020a,TESS_QLP_Huang2020b,TESS_QLP_Fausnaugh2020})
for FGK stars with $G < 12$~mag that were observed in Sector 5 (chosen arbitrarily). 
For each star, we used the \texttt{lightkurve}
package \citep{Lightkurve18} to compute the Combined Differential Photometric
Precision (CDPP; \citealt{Christiansen2012,vanCleve2016})
on a 2-hour timescale.
The transit detection probability is chiefly
a function of CDPP.

\begin{figure}
\plotone{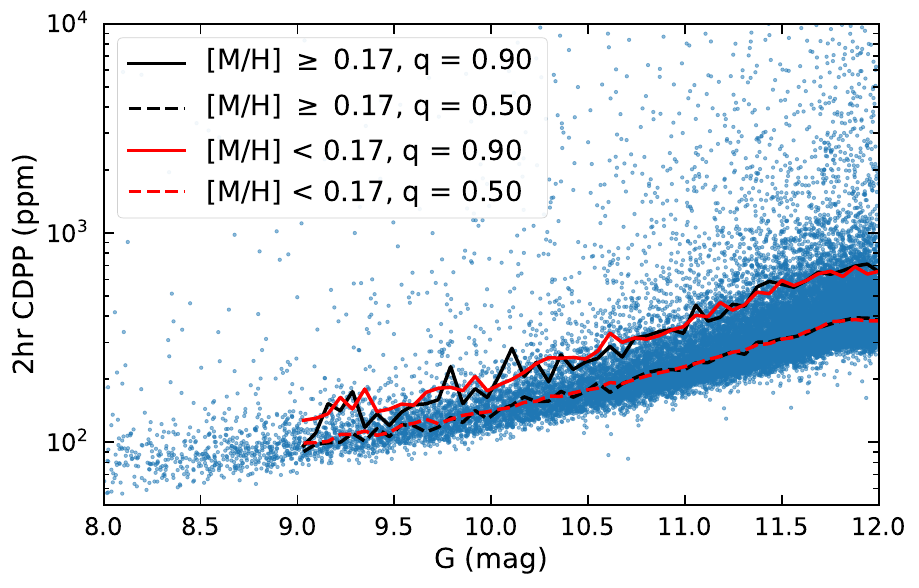}
\caption{Estimated 2-hr CDPP values for FGK stars brighter than $G = 12$~mag
observed by TESS in Sector 5 (blue dots), as a function of apparent magnitude.
The solid and dashed lines show the 90th and 50th percentiles
in magnitude bins of width 0.05. The black and red lines represent
high and low metallicity subsamples, respectively,
divided at the median metallicity of the hot Jupiter host sample.
\label{fig:cdpp_dist}}
\end{figure}

Figure \ref{fig:cdpp_dist} shows the CDPP as a function
of Gaia apparent magnitude. Also shown are CDPP quantiles
for the metal-poor and metal-rich stars separately, based on
the calibrated \Gaia GSP-Spec metallicities.
We found no significant difference in the photometric
noise properties of the two subsamples. This eliminates
the possibility that metal-rich stars have systematically noisier light
curves than metal-poor stars, or vice versa.

Despite this reassurance, there may be other sources of metallicity-dependent period biases. Using a more complete sample of planets
would reduce the possibility that a large number of undetected planets
exist that have a different metallicity/period distribution.
Constructing a magnitude-limited sample of hot Jupiters is
the goal of the Grand Unified Hot Jupiter Survey
\citep{Yee2022a,Yee2022b}.
Although this goal has not yet been achieved,
completeness has been substantially increased over
the last few years.
There are also many hot Jupiters
for which confirmation
has been completed but 
has only been described in preprints or works in preparation,
and are therefore not included in the NEA.

We decided to augment the NEA-based sample with 16 hot Jupiters
described only in preprints \citep{Anderson2014,Anderson2018,Bakos2016,Brown2019,
Rodriguez2022,Sha2022,Yee2022b} at the time of archive query, and 22 that
will be described in forthcoming papers
by Schulte et al., Yee et al., and Quinn et al.\ that meet the
same criteria on planet and stellar properties described in \S\ref{sec:exoarchive}.
This brought the total sample size to 240.
Based on the simulations in
\citet{Yee2021b}, we expect that this enlarged sample is $\approx$80\%
complete to $G=12$, and therefore that any biases are limited
to the 10--20\% level.

\Gaia\ \gspspec
metallicities are available for 232 (97\%) of the host stars in the sample.
After performing the correction described in Section~\ref{ssec:gaia_metallicities}, the median metallicity is
$+0.171$~dex, consistent with previous homogeneous studies of giant-planet hosts
\citep{Buchhave2018,Petigura2018}.
Thus, the high completeness, high
reliability, and homogeneous metallicity scale
for this sample improves over the previous works by \citet{Dawson2013} and \citet{Hellier2017}.

The right panel of
Figure \ref{fig:period_dist} displays the period distributions
of the high-metallicity and low-metallicity halves of the sample,
which appear to be similar.
A K-S test gives a $p$-value of 0.46.
When the sample is split at $\mh=0$ instead of the median
value of $+0.171$, the $p$-value is 0.64.
In neither case can we reject the hypothesis that the periods in each subsample are drawn from the same distribution.

\section{Kepler Sample} \label{sec:kepler_hjs}

\begin{figure}
\plotone{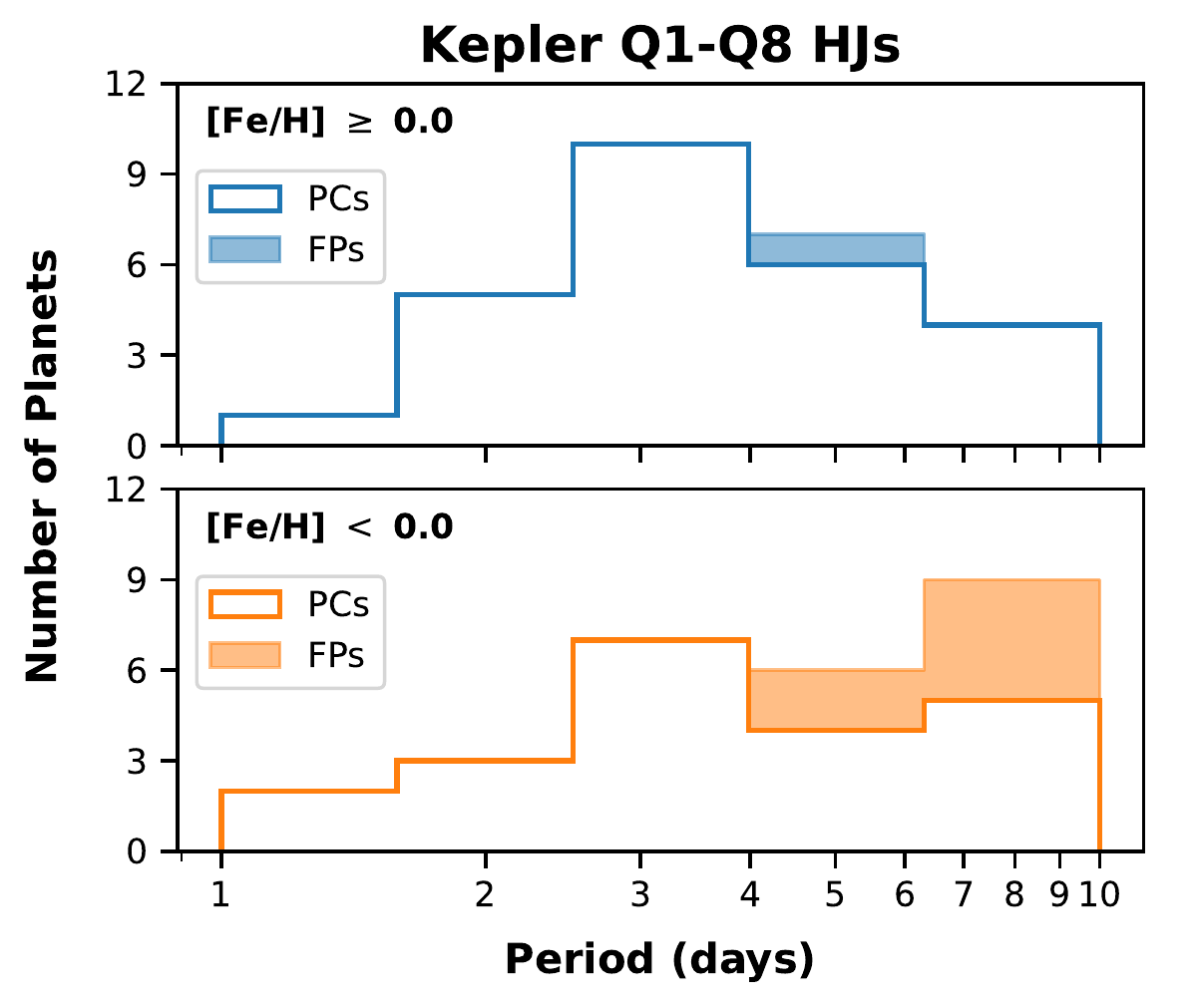}
\caption{
Period distributions of hot-Jupiter
candidates
from the first eight quarters of the \Kepler mission (the same catalog used
by \citet{Dawson2013}). Within the $P < 10$~day range, the period distribution
for planets orbiting metal-rich stars (top) appears  distinct from that
for metal-poor stars (bottom), but the number of objects is too small to draw
any statistical conclusions (K-S test $p$-value = 0.24).
Furthermore, when excluding known false positives (FPs, shaded region), leaving only
planet candidates (PCs), the differences between the two distributions are reduced.
\label{fig:kepler_hjs_fps}}
\end{figure}

With our larger and purer sample of hot Jupiters, we did not find any
evidence that their orbital period distribution depends on the metallicity
of their host stars.
We sought to understand if this finding is discrepant with the period distribution
of \Kepler hot Jupiters.

\citet{Dawson2013} found a significant difference
when comparing the period distributions of all
the \Kepler\ giant-planet candidates, split into metal-rich and metal-poor samples.
This is in contrast to our
comparisons,
which were restricted
to $P<10$~days.
Nonetheless, it is worth investigating if our result is corroborated in the
\Kepler data, which has the advantage of being from a single homogeneous survey
that would have detected essentially all the transiting short-period
giant planets orbiting FGK stars for which data were collected.

We downloaded the
list of planet candidates from the first eight quarters of the \Kepler mission
\citep{Burke2014},the same catalog examined in \citet{Dawson2013}.
We made the same cuts on planet and stellar host properties as in \S\ref{sec:exoarchive},
restricted the sample to $P < 10$~days,
and split the sample according to the metallicities provided in the Kepler Input
Catalog (KIC; \citealt{Brown2011})
, using a boundary of $\feh = 0.0$.
Figure \ref{fig:kepler_hjs_fps} shows the period distribution of these two
subsamples. The period distribution for planets orbiting metal-rich stars
appears to have a peak at $\sim3$~days, while the number of planets
orbiting metal-poor stars increases steadily out to $P = 10$~days.

However, even though \Kepler searched $\sim10^5$ stars, the intrinsic occurrence rate
of hot Jupiters is low, resulting in a small sample.
The full Q1-Q8 \Kepler sample contains only 56 hot Jupiter candidates.
A K-S test comparing only the \Kepler hot Jupiters
gives a $p$-value
of 0.24, too large to conclude that the period distributions differ
between metal-poor and metal-rich hosts.

Furthermore, most of the stars that were searched by the
\Kepler mission are too faint for detailed follow-up and
characterization of planet candidates.
Even today, most of the planets remain only as ``validated''
(deemed likely to be planets based on statistical considerations)
as opposed to ``confirmed'' by Doppler spectroscopy.
Of the 56 hot Jupiter candidates, seven were subsequently determined to be false
positives, and five others were still unvalidated candidates as of the end of the mission
\citep{Twicken2016}. If we remove the known false positives from the orbital
period distribution (shaded histogram in Figure \ref{fig:kepler_hjs_fps}), the
differences between the two subsamples diminish further.
Thus, our results do not contradict the results of the \Kepler survey.

\section{Discussion}

Does the lack of any detectable difference between the orbital period
distributions of hot Jupiters orbiting metal-rich stars and metal-poor stars
allow us to place any constraints on the relative fraction of hot Jupiters
that formed in different ways? Since we are not aware of any
theory that provides specific quantitative predictions for the period distribution, we answered this question using a relatively simple parametric model inspired by the work of \citet{Ford2006}
and \citet{Nelson2017}.

If hot Jupiters migrate to their current orbital positions through
high-eccentricity migration, the minimum periastron distance they can reach
without becoming tidally disrupted is the Roche distance, \aroche.
Planetary orbits that initially have large semimajor axes (large enough
to extend beyond the ``ice line'' where giant planet formation is expected)
and a periastron distance of $\aroche$ will end up on circular orbits
of radius $2\aroche$, after tidal circularization and assuming no loss
of angular momentum. Thus, in this scenario, we expect the distribution
of orbital distances to cut off below $2\aroche$, corresponding
to a minimum orbital period as a function of the planet's mean density
\citep{Rappaport2013}.
In contrast, if hot Jupiters arise from disk-driven migration,
they may be able to migrate all the way to a circular orbit of radius
$\aroche$.

Based on this reasoning, \citet{Nelson2017} modeled the giant planet population using
a two-component truncated power law model, $dN/dx \propto x^{\gamma-1}$, where
$x \equiv a/\aroche$ and
the lower truncation limit is $x=1$ for disk-driven migration
and $x=2$ for high-eccentricity migration.
They found that the data could be explained if the
disk-driven migration component (Pop.~1) follows a shallow power law,
and the high-eccentricity migration component (Pop.~2) follows a steeper power law.
They found the fraction of planets belonging to Pop.~1 to be
0.15--0.35, depending on the sample
used. They also found that such a model was preferred over a single-
component power law.

\begin{figure}
\plotone{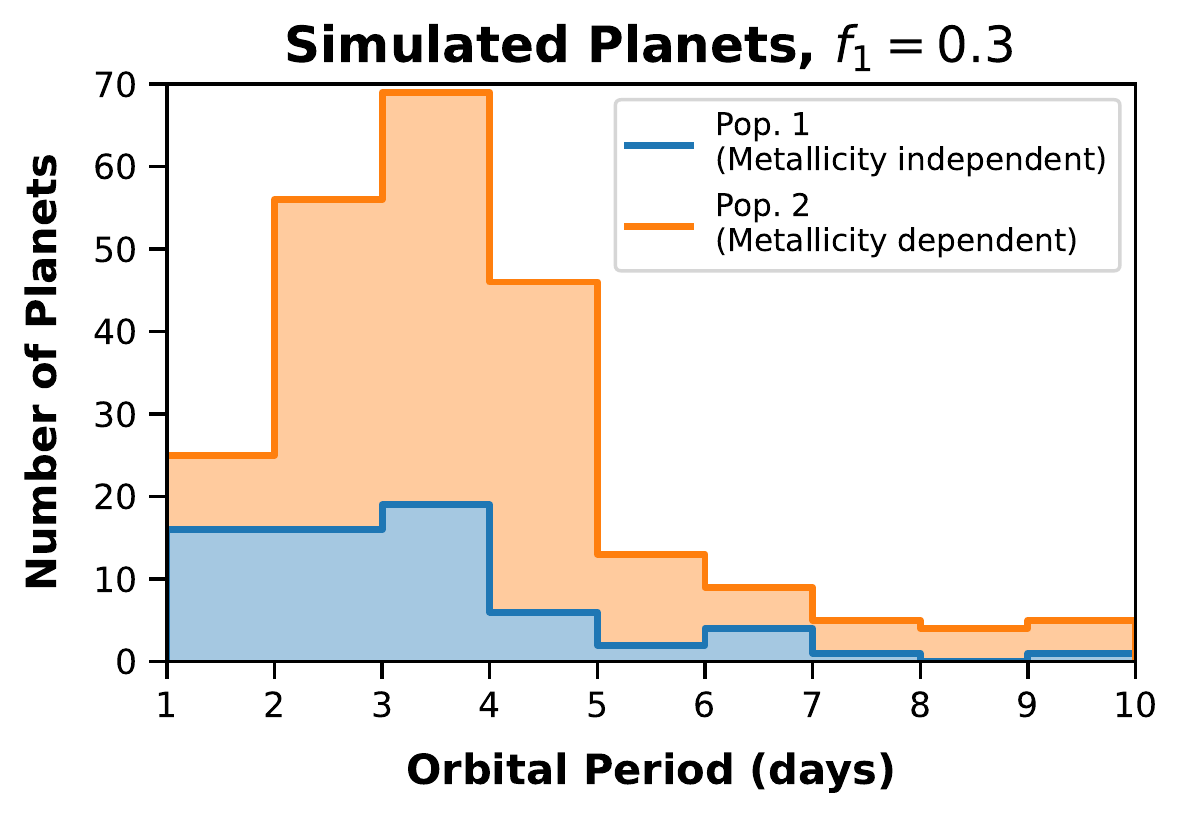}
\plotone{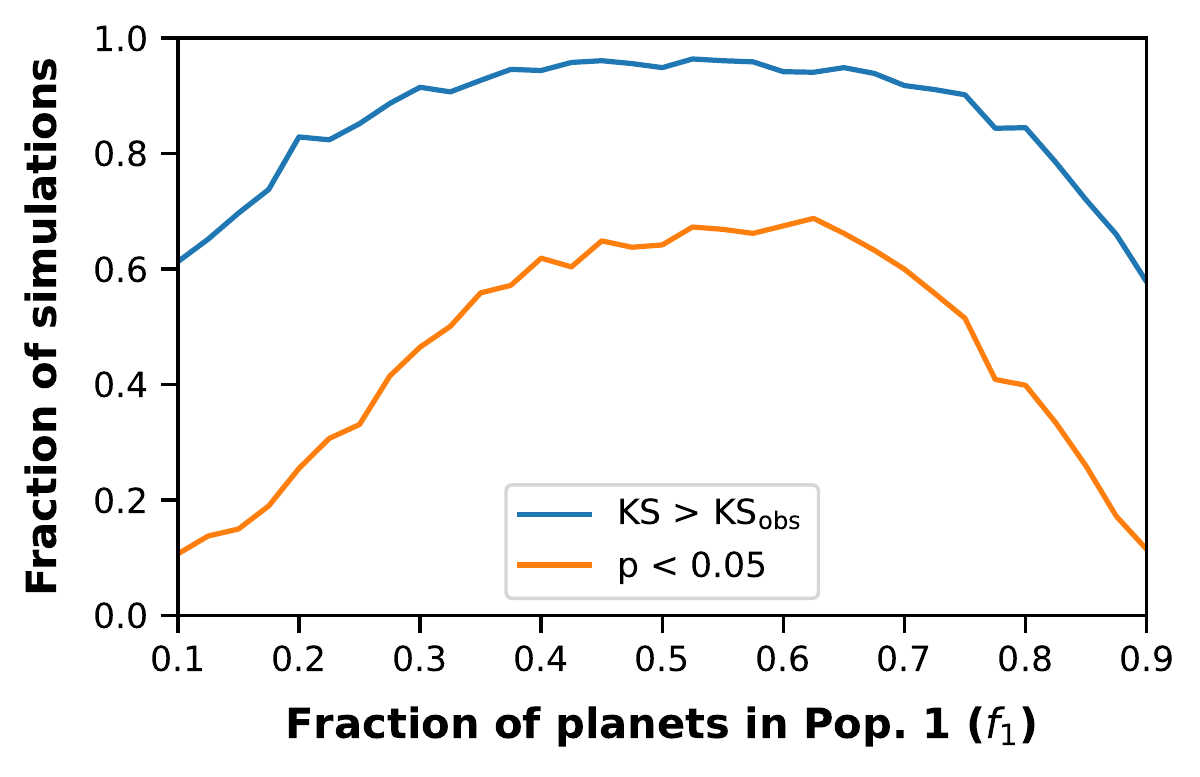}
\caption{\textbf{Top:} Example Monte Carlo sample of planets.
The blue histogram is for the disk-driven migration component,
obtained by drawing $x \equiv a/\aroche$ from a shallow power-law distribution 
($\gamma_1 = -0.04$) truncated
at $x=1$., blue histogram)
The orange histogram is for the high-eccentricity migration component,
obtained by drawing $x$ from a steeper power-law distribution($\gamma_2 = -1.38$)
truncated at $x=2$. Both distributions were further attenuated by
the transit probability, $\Rstar/a$.
\textbf{Bottom:} Blue line: The fraction of simulations where the K-S statistic
is larger than that for the observed hot Jupiters.
Orange line: The fraction of simulations in which the null hypothesis
that the period distribution is the same for high- and low-metallicity stars
could be ruled out by the K-S test with $>$95\% confidence.
\label{fig:sim_hj_pop}}
\end{figure}

In this picture, one might expect the occurrence of Pop.~1 planets
to have a relatively weak dependence on metallicity,
while the occurrence of Pop.~2 planets
is higher for metal-rich stars that can form multiple giant
planets that eventually undergo scattering events. We simulated both
populations of planets, sampling $x$ from power law distributions with
parameters fixed at the best-fit values from \citet{Nelson2017}
($x_{l,1} = 1, \gamma_1 = -0.04, x_{l,2} = 2, \gamma_2 = =1.38$),
and planet masses and radii drawn from those of the known hot Jupiters. We then
drew random samples of 240 planets each from the two populations, varying
the relative fraction $f_1$ of planets in Pop.~1 between 0.1 and 0.5,
and accounting for transit
probability (Figure \ref{fig:sim_hj_pop}, top panel).
We assigned planets from Pop.~1 to stars brighter than $G < 12$
from the \Gaia DR3 catalog \citep{GaiaDR3Summary_GaiaCollaboration2022a}
independently of metallicity. Meanwhile, planets from Pop.~2 were assigned
to stars with probability $\mathrm{Pr.} \propto 10^{\beta \mathrm{\feh}}$, 
the exponential relation seen
for the overall hot Jupiter population \citep[e.g.,][]{Valenti2005}.
We chose $\beta = 3.4$ based on the study of \Kepler planet occurrence by
\citet{Petigura2018}.

Using these simulated planet samples, we performed the same K-S tests that we performed
on the
actual sample of planets, and repeated this process 1{,}000 times for each
choice of the relative fraction $f_1$.
As $f_1$ is increased, the peak of the orbital period distribution in the metal-poor subsample shifts to smaller orbital periods.
The lower panel of Figure \ref{fig:sim_hj_pop} shows the
fraction of simulations 
in which the low-metallicity and high-metallicity subsamples had
differences in orbital period distribution larger than that seen in the observed planet sample as measured by the K-S statistic.
We also counted the fraction of trials in which the K-S test could reject with 95\%
confidence the null hypothesis that the metal-rich and metal-poor subsamples are drawn from the same distribution.
When the sample comprises a roughly equal admixture of both populations ($0.35 \lesssim f_1 \lesssim 0.75$), the null hypothesis was rejected in the majority of simulations.
When the relative fraction of either population is smaller ($f_1 \lesssim 0.3$ or $f_1 \gtrsim 0.8$), there is no longer a significant difference in the period distributions of the two subsamples.

Thus, in the context of our adopted model, the absence of a detectable difference
between the period distributions of hot Jupiters orbiting metal-rich and metal-poor stars
rules out 
$0.35 \lesssim f_1 \lesssim 0.75$, and allows for smaller relative fractions of either population.
However, we stress that
this analysis is based on a simple parametric model in which most of the constraining power
arises from the shortest period planets; more detailed theoretical predictions
of the form of the orbital period distribution arising from these two migration
pathways would likely be more informative, even when applied to the same sample.

\section{Conclusion}

Studies of the metallicity distribution of planet-hosting stars
is a promising method for 
distinguishing planet populations
that formed in different ways \citep[see, e.g.,][]{Winn2017,Schlaufman2018,Buchhave2018}.
We revisited the possibility that the orbital period distribution
of hot Jupiters depends on their host stars' metallicities, but found
no evidence for such an effect. 
This corroborates the results of \citet{Hellier2017} with a more homogeneous and complete dataset.

\citet{Dawson2013}
used the \Kepler\ data to argue that the short-period pile-up of giant planets
is a feature only of metal-rich host stars.
The lack of difference in period distribution for planets
within the $P < 10$~day range weakens that line of argument.
Nonetheless, \citet{Dawson2013}
presented two other lines of evidence for this formation pathway: that
gas giants orbiting metal-rich stars span a range of eccentricities in
contrast to those orbiting metal-poor stars, and that most eccentric
proto-hot Jupiters appear to be orbiting metal-rich stars. These were
based on studies of planets found by the radial-velocity and ground-based
transit surveys, and are unaffected by our work.

The lack of a detectable metallicity dependence of
the period distribution may itself be a clue about the formation of hot Jupiters.
Within the context of the two-pathway model of
\citet{Nelson2017} --- in which a metallicity-dependent formation
pathway brings hot Jupiters close
to the Roche limit, while a metallicity-indepedent
pathway delivers hot Jupiters to distances
no closer than twice the Roche limit ---
the data imply that
one of these pathways must
produce $\gtrsim 70\%$ of the population.
More quantitative
theoretical work to predict the period distribution from different
formation pathways \citep[e.g.,][]{Heller2019} would be required to
place more reliable constraints. Alternatively, jointly modeling the
orbital period and eccentricity distributions may help us identify
the planets arising from each pathway.

Thanks to the success of all-sky survey missions such as TESS and \textit{Gaia},
hot Jupiter demographics are coming into sharper focus.
Previous work was based on heterogeneous samples from
ground-based transit surveys, or the smaller number of planets drawn from
\Kepler or radial-velocity searches. The combination of TESS and \textit{Gaia}
is allowing a much larger statistical sample to be assembled.
Combining a large sample of true planets with an assessment
of the detection efficiency of TESS will also enable a measurement
of occurrence rates \citep{Zhou2019a,Beleznay2022} and their
intrinsic period distribution.
\newline

\begin{acknowledgements}
We thank the anonymous referee for helpful suggestions which helped improve and clarify this manuscriipt.
This research made use of the NASA Exoplanet Archive, which is operated by the California Institute of Technology, under contract with the National Aeronautics and Space Administration under the Exoplanet Exploration Program.
This research also made use of data from the European Space Agency (ESA) mission
{\it Gaia} (\url{https://www.cosmos.esa.int/gaia}), processed by the {\it Gaia}
Data Processing and Analysis Consortium (DPAC,
\url{https://www.cosmos.esa.int/web/gaia/dpac/consortium}). Funding for the DPAC
has been provided by national institutions, in particular the institutions
participating in the {\it Gaia} Multilateral Agreement.
\end{acknowledgements}

\bibliography{manuscript}

\begin{thebibliography}{}
\expandafter\ifx\csname natexlab\endcsname\relax\def\natexlab#1{#1}\fi
\providecommand{\url}[1]{\href{#1}{#1}}
\providecommand{\dodoi}[1]{doi:~\href{http://doi.org/#1}{\nolinkurl{#1}}}
\providecommand{\doeprint}[1]{\href{http://ascl.net/#1}{\nolinkurl{http://ascl.net/#1}}}
\providecommand{\doarXiv}[1]{\href{https://arxiv.org/abs/#1}{\nolinkurl{https://arxiv.org/abs/#1}}}

\bibitem[{Anderson {et~al.}(2014)Anderson, Brown, Cameron, Delrez, Fumel,
  Gillon, Hellier, Jehin, Lendl, Maxted, {Neveu-VanMalle}, Pepe, Pollacco,
  Queloz, Rojo, Segransan, Serenelli, Smalley, Smith, Southworth, Triaud,
  Turner, Udry, \& West}]{Anderson2014}
Anderson, D.~R., Brown, D. J.~A., Cameron, A.~C., {et~al.} 2014, Six
  Newly-Discovered Hot {{Jupiters}} Transiting {{F}}/{{G}} Stars: {{WASP-87b}},
  {{WASP-108b}}, {{WASP-109b}}, {{WASP-110b}}, {{WASP-111b}} \& {{WASP-112b}},
  {arXiv}.
\newblock \doarXiv{1410.3449}

\bibitem[{Anderson {et~al.}(2018)Anderson, Bouchy, Brown, Cameron, Delrez,
  Gillon, Hern{\'a}ndez, Hellier, Jehin, Lendl, Maxted, {Neveu-VanMalle},
  Nielsen, Pepe, Perger, Pollacco, Queloz, Rey, S{\'e}gransan, Smalley,
  {Toledo-Padr{\'o}n}, Triaud, Turner, Udry, \& West}]{Anderson2018}
Anderson, D.~R., Bouchy, F., Brown, D. J.~A., {et~al.} 2018, The Discovery of
  {{WASP-134b}}, {{WASP-134c}}, {{WASP-137b}}, {{WASP-143b}} and {{WASP-146b}}:
  Three Hot {{Jupiters}} and a Pair of Warm {{Jupiters}} Orbiting
  {{Solar-type}} Stars,  {arXiv}, \dodoi{10.48550/arXiv.1812.09264}

\bibitem[{Bakos {et~al.}(2016)Bakos, Hartman, Torres, Latham, Sato, Bieryla,
  Shporer, Howard, Fulton, Buchhave, Penev, Kov{\'a}cs, Kov{\'a}cs, Csubry,
  Esquerdo, Everett, Szklen{\'a}r, Quinn, B{\'e}ky, Marcy, Noyes,
  L{\'a}z{\'a}r, Papp, \& S{\'a}ri}]{Bakos2016}
Bakos, G.~{\'A}., Hartman, J.~D., Torres, G., {et~al.} 2016, {{HAT-P-47b AND
  HAT-P-48b}}: {{Two Low Density Sub-Saturn-Mass Transiting Planets}} on the
  {{Edge}} of the {{Period--Mass Desert}},  {arXiv},
  \dodoi{10.48550/arXiv.1606.04556}

\bibitem[{Beleznay \& Kunimoto(2022)}]{Beleznay2022}
Beleznay, M., \& Kunimoto, M. 2022, Monthly Notices of the Royal Astronomical
  Society, 516, 75, \dodoi{10.1093/mnras/stac2179}

\bibitem[{Brewer {et~al.}(2016)Brewer, Fischer, Valenti, \&
  Piskunov}]{Brewer2016}
Brewer, J.~M., Fischer, D.~A., Valenti, J.~A., \& Piskunov, N. 2016, The
  Astrophysical Journal Supplement Series, 225, 32,
  \dodoi{10.3847/0067-0049/225/2/32}

\bibitem[{Brown {et~al.}(2019)Brown, Anderson, Doyle, Maxted, Smalley,
  McCormac, Almenera, {Prieto-Arranz}, Deleuil, Diaz, Foxell, Hebrard, Lendl,
  Delrez, Gillon, Jehin, Lam, Triaud, Turner, Armstrong, Bouchy, Cameron,
  Pollacco, Faedi, Chew, Hebb, Hellier, {Neveu-VanMalle}, Palle, Queloz,
  Segransan, Udry, \& West}]{Brown2019}
Brown, D. J.~A., Anderson, D.~R., Doyle, A.~P., {et~al.} 2019, Three Transiting
  Planet Discoveries from the {{Wide Angle Search}} for {{Planets}}: {{WASP-85
  A}} b; {{WASP-116}} b, and {{WASP-149}} b,  {arXiv}.
\newblock \doarXiv{1412.7761}

\bibitem[{Brown {et~al.}(2011)Brown, Latham, Everett, \& Esquerdo}]{Brown2011}
Brown, T.~M., Latham, D.~W., Everett, M.~E., \& Esquerdo, G.~A. 2011, The
  Astronomical Journal, 142, 112, \dodoi{10.1088/0004-6256/142/4/112}

\bibitem[{Buchhave {et~al.}(2018)Buchhave, Bitsch, Johansen, Latham, Bizzarro,
  Bieryla, \& Kipping}]{Buchhave2018}
Buchhave, L.~A., Bitsch, B., Johansen, A., {et~al.} 2018, The Astrophysical
  Journal, 856, 37, \dodoi{10.3847/1538-4357/aaafca}

\bibitem[{Burke {et~al.}(2014)Burke, Bryson, Mullally, Rowe, Christiansen,
  Thompson, Coughlin, Haas, Batalha, Caldwell, Jenkins, Still, Barclay,
  Borucki, Chaplin, Ciardi, Clarke, Cochran, Demory, Esquerdo, Gautier,
  Gilliland, Girouard, Havel, Henze, Howell, Huber, Latham, Li, Morehead,
  Morton, Pepper, Quintana, Ragozzine, Seader, Shah, Shporer, Tenenbaum,
  Twicken, \& Wolfgang}]{Burke2014}
Burke, C.~J., Bryson, S.~T., Mullally, F., {et~al.} 2014, The Astrophysical
  Journal Supplement Series, 210, 19, \dodoi{10.1088/0067-0049/210/2/19}

\bibitem[{Butler {et~al.}(2006)Butler, Wright, Marcy, Fischer, Vogt, Tinney,
  Jones, Carter, Johnson, McCarthy, \& Penny}]{Butler2006}
Butler, R.~P., Wright, J.~T., Marcy, G.~W., {et~al.} 2006, The Astrophysical
  Journal, 646, 505, \dodoi{10.1086/504701}

\bibitem[{Christiansen {et~al.}(2012)Christiansen, Jenkins, Barclay, Burke,
  Caldwell, Clarke, Li, Seader, Smith, Stumpe, Tenenbaum, Thompson, Twicken, \&
  Van~Cleve}]{Christiansen2012}
Christiansen, J.~L., Jenkins, J.~M., Barclay, T.~S., {et~al.} 2012,
  Publications of the Astronomical Society of the Pacific, 124, 1279,
  \dodoi{10.1086/668847}

\bibitem[{Cropper {et~al.}(2018)Cropper, Katz, Sartoretti, Prusti, de~Bruijne,
  Chassat, Charvet, Boyadjian, Perryman, Sarri, Gare, Erdmann, Munari, Zwitter,
  Wilkinson, Arenou, Vallenari, G{\'o}mez, Panuzzo, Seabroke, Prieto, Benson,
  Marchal, Huckle, Smith, Dolding, Jan{\ss}en, Viala, Blomme, Baker,
  Boudreault, Crifo, Soubiran, Fr{\'e}mat, Jasniewicz, Guerrier, Guy, Turon,
  {Jean-Antoine-Piccolo}, Th{\'e}venin, David, Gosset, \&
  Damerdji}]{GaiaRVS_Cropper2018}
Cropper, M., Katz, D., Sartoretti, P., {et~al.} 2018, Astronomy \&
  Astrophysics, 616, A5, \dodoi{10.1051/0004-6361/201832763}

\bibitem[{{Cumming} {et~al.}(2008){Cumming}, {Butler}, {Marcy}, {Vogt},
  {Wright}, \& {Fischer}}]{Cumming2008}
{Cumming}, A., {Butler}, R.~P., {Marcy}, G.~W., {et~al.} 2008, \pasp, 120, 531,
  \dodoi{10.1086/588487}

\bibitem[{Dawson \& {Murray-Clay}(2013)}]{Dawson2013}
Dawson, R.~I., \& {Murray-Clay}, R.~A. 2013, The Astrophysical Journal Letters,
  6

\bibitem[{Fausnaugh {et~al.}(2020)Fausnaugh, Burke, Ricker, \&
  Vanderspek}]{TESS_QLP_Fausnaugh2020}
Fausnaugh, M.~M., Burke, C.~J., Ricker, G.~R., \& Vanderspek, R. 2020, Research
  Notes of the AAS, 4, 251, \dodoi{10.3847/2515-5172/abd63a}

\bibitem[{Fischer \& Valenti(2005)}]{Fischer2005}
Fischer, D.~A., \& Valenti, J. 2005, The Astrophysical Journal, 622, 1102,
  \dodoi{10.1086/428383}

\bibitem[{Ford \& Rasio(2006)}]{Ford2006}
Ford, E.~B., \& Rasio, F.~A. 2006, The Astrophysical Journal Letters, 638, L45,
  \dodoi{10.1086/500734}

\bibitem[{{Gaia Collaboration} {et~al.}(2022){Gaia Collaboration}, Vallenari,
  Brown, Prusti, \& {et al.}}]{GaiaDR3Summary_GaiaCollaboration2022a}
{Gaia Collaboration}, Vallenari, A., Brown, A., Prusti, T., \& {et al.} 2022,
  Astronomy \& Astrophysics, \dodoi{10.1051/0004-6361/202243940}

\bibitem[{{Gaia Collaboration} {et~al.}(2016){Gaia Collaboration}, {Prusti},
  {de Bruijne}, {Brown}, {Vallenari}, {Babusiaux}, {Bailer-Jones}, {Bastian},
  {Biermann}, {Evans}, {Eyer}, {Jansen}, {Jordi}, {Klioner}, {Lammers},
  {Lindegren}, {Luri}, {Mignard}, {Milligan}, {Panem}, {Poinsignon},
  {Pourbaix}, {Randich}, {Sarri}, {Sartoretti}, {Siddiqui}, {Soubiran},
  {Valette}, {van Leeuwen}, {Walton}, {Aerts}, {Arenou}, {Cropper}, {Drimmel},
  {H{\o}g}, {Katz}, {Lattanzi}, {O'Mullane}, {Grebel}, {Holland}, {Huc},
  {Passot}, {Bramante}, {Cacciari}, {Casta{\~n}eda}, {Chaoul}, {Cheek}, {De
  Angeli}, {Fabricius}, {Guerra}, {Hern{\'a}ndez}, {Jean-Antoine-Piccolo},
  {Masana}, {Messineo}, {Mowlavi}, {Nienartowicz}, {Ord{\'o}{\~n}ez-Blanco},
  {Panuzzo}, {Portell}, {Richards}, {Riello}, {Seabroke}, {Tanga},
  {Th{\'e}venin}, {Torra}, {Els}, {Gracia-Abril}, {Comoretto},
  {Garcia-Reinaldos}, {Lock}, {Mercier}, {Altmann}, {Andrae}, {Astraatmadja},
  {Bellas-Velidis}, {Benson}, {Berthier}, {Blomme}, {Busso}, {Carry},
  {Cellino}, {Clementini}, {Cowell}, {Creevey}, {Cuypers}, {Davidson}, {De
  Ridder}, {de Torres}, {Delchambre}, {Dell'Oro}, {Ducourant}, {Fr{\'e}mat},
  {Garc{\'\i}a-Torres}, {Gosset}, {Halbwachs}, {Hambly}, {Harrison}, {Hauser},
  {Hestroffer}, {Hodgkin}, {Huckle}, {Hutton}, {Jasniewicz}, {Jordan},
  {Kontizas}, {Korn}, {Lanzafame}, {Manteiga}, {Moitinho}, {Muinonen},
  {Osinde}, {Pancino}, {Pauwels}, {Petit}, {Recio-Blanco}, {Robin}, {Sarro},
  {Siopis}, {Smith}, {Smith}, {Sozzetti}, {Thuillot}, {van Reeven}, {Viala},
  {Abbas}, {Abreu Aramburu}, {Accart}, {Aguado}, {Allan}, {Allasia},
  {Altavilla}, {{\'A}lvarez}, {Alves}, {Anderson}, {Andrei}, {Anglada Varela},
  {Antiche}, {Antoja}, {Ant{\'o}n}, {Arcay}, {Atzei}, {Ayache}, {Bach},
  {Baker}, {Balaguer-N{\'u}{\~n}ez}, {Barache}, {Barata}, {Barbier}, {Barblan},
  {Baroni}, {Barrado y Navascu{\'e}s}, {Barros}, {Barstow}, {Becciani},
  {Bellazzini}, {Bellei}, {Bello Garc{\'\i}a}, {Belokurov}, {Bendjoya},
  {Berihuete}, {Bianchi}, {Bienaym{\'e}}, {Billebaud}, {Blagorodnova},
  {Blanco-Cuaresma}, {Boch}, {Bombrun}, {Borrachero}, {Bouquillon}, {Bourda},
  {Bouy}, {Bragaglia}, {Breddels}, {Brouillet}, {Br{\"u}semeister},
  {Bucciarelli}, {Budnik}, {Burgess}, {Burgon}, {Burlacu}, {Busonero}, {Buzzi},
  {Caffau}, {Cambras}, {Campbell}, {Cancelliere}, {Cantat-Gaudin}, {Carlucci},
  {Carrasco}, {Castellani}, {Charlot}, {Charnas}, {Charvet}, {Chassat},
  {Chiavassa}, {Clotet}, {Cocozza}, {Collins}, {Collins}, {Costigan}, {Crifo},
  {Cross}, {Crosta}, {Crowley}, {Dafonte}, {Damerdji}, {Dapergolas}, {David},
  {David}, {De Cat}, {de Felice}, {de Laverny}, {De Luise}, {De March}, {de
  Martino}, {de Souza}, {Debosscher}, {del Pozo}, {Delbo}, {Delgado},
  {Delgado}, {di Marco}, {Di Matteo}, {Diakite}, {Distefano}, {Dolding}, {Dos
  Anjos}, {Drazinos}, {Dur{\'a}n}, {Dzigan}, {Ecale}, {Edvardsson}, {Enke},
  {Erdmann}, {Escolar}, {Espina}, {Evans}, {Eynard Bontemps}, {Fabre},
  {Fabrizio}, {Faigler}, {Falc{\~a}o}, {Farr{\`a}s Casas}, {Faye}, {Federici},
  {Fedorets}, {Fern{\'a}ndez-Hern{\'a}ndez}, {Fernique}, {Fienga}, {Figueras},
  {Filippi}, {Findeisen}, {Fonti}, {Fouesneau}, {Fraile}, {Fraser}, {Fuchs},
  {Furnell}, {Gai}, {Galleti}, {Galluccio}, {Garabato}, {Garc{\'\i}a-Sedano},
  {Gar{\'e}}, {Garofalo}, {Garralda}, {Gavras}, {Gerssen}, {Geyer}, {Gilmore},
  {Girona}, {Giuffrida}, {Gomes}, {Gonz{\'a}lez-Marcos},
  {Gonz{\'a}lez-N{\'u}{\~n}ez}, {Gonz{\'a}lez-Vidal}, {Granvik}, {Guerrier},
  {Guillout}, {Guiraud}, {G{\'u}rpide}, {Guti{\'e}rrez-S{\'a}nchez}, {Guy},
  {Haigron}, {Hatzidimitriou}, {Haywood}, {Heiter}, {Helmi}, {Hobbs},
  {Hofmann}, {Holl}, {Holland}, {Hunt}, {Hypki}, {Icardi}, {Irwin}, {Jevardat
  de Fombelle}, {Jofr{\'e}}, {Jonker}, {Jorissen}, {Julbe}, {Karampelas},
  {Kochoska}, {Kohley}, {Kolenberg}, {Kontizas}, {Koposov}, {Kordopatis},
  {Koubsky}, {Kowalczyk}, {Krone-Martins}, {Kudryashova}, {Kull}, {Bachchan},
  {Lacoste-Seris}, {Lanza}, {Lavigne}, {Le Poncin-Lafitte}, {Lebreton},
  {Lebzelter}, {Leccia}, {Leclerc}, {Lecoeur-Taibi}, {Lemaitre}, {Lenhardt},
  {Leroux}, {Liao}, {Licata}, {Lindstr{\o}m}, {Lister}, {Livanou}, {Lobel},
  {L{\"o}ffler}, {L{\'o}pez}, {Lopez-Lozano}, {Lorenz}, {Loureiro},
  {MacDonald}, {Magalh{\~a}es Fernandes}, {Managau}, {Mann}, {Mantelet},
  {Marchal}, {Marchant}, {Marconi}, {Marie}, {Marinoni}, {Marrese},
  {Marschalk{\'o}}, {Marshall}, {Mart{\'\i}n-Fleitas}, {Martino}, {Mary},
  {Matijevi{\v{c}}}, {Mazeh}, {McMillan}, {Messina}, {Mestre}, {Michalik},
  {Millar}, {Miranda}, {Molina}, {Molinaro}, {Molinaro}, {Moln{\'a}r},
  {Moniez}, {Montegriffo}, {Monteiro}, {Mor}, {Mora}, {Morbidelli}, {Morel},
  {Morgenthaler}, {Morley}, {Morris}, {Mulone}, {Muraveva}, {Musella},
  {Narbonne}, {Nelemans}, {Nicastro}, {Noval}, {Ord{\'e}novic},
  {Ordieres-Mer{\'e}}, {Osborne}, {Pagani}, {Pagano}, {Pailler}, {Palacin},
  {Palaversa}, {Parsons}, {Paulsen}, {Pecoraro}, {Pedrosa}, {Pentik{\"a}inen},
  {Pereira}, {Pichon}, {Piersimoni}, {Pineau}, {Plachy}, {Plum}, {Poujoulet},
  {Pr{\v{s}}a}, {Pulone}, {Ragaini}, {Rago}, {Rambaux}, {Ramos-Lerate},
  {Ranalli}, {Rauw}, {Read}, {Regibo}, {Renk}, {Reyl{\'e}}, {Ribeiro},
  {Rimoldini}, {Ripepi}, {Riva}, {Rixon}, {Roelens}, {Romero-G{\'o}mez},
  {Rowell}, {Royer}, {Rudolph}, {Ruiz-Dern}, {Sadowski}, {Sagrist{\`a}
  Sell{\'e}s}, {Sahlmann}, {Salgado}, {Salguero}, {Sarasso}, {Savietto},
  {Schnorhk}, {Schultheis}, {Sciacca}, {Segol}, {Segovia}, {Segransan},
  {Serpell}, {Shih}, {Smareglia}, {Smart}, {Smith}, {Solano}, {Solitro},
  {Sordo}, {Soria Nieto}, {Souchay}, {Spagna}, {Spoto}, {Stampa}, {Steele},
  {Steidelm{\"u}ller}, {Stephenson}, {Stoev}, {Suess}, {S{\"u}veges}, {Surdej},
  {Szabados}, {Szegedi-Elek}, {Tapiador}, {Taris}, {Tauran}, {Taylor},
  {Teixeira}, {Terrett}, {Tingley}, {Trager}, {Turon}, {Ulla}, {Utrilla},
  {Valentini}, {van Elteren}, {Van Hemelryck}, {van Leeuwen}, {Varadi},
  {Vecchiato}, {Veljanoski}, {Via}, {Vicente}, {Vogt}, {Voss}, {Votruba},
  {Voutsinas}, {Walmsley}, {Weiler}, {Weingrill}, {Werner}, {Wevers},
  {Whitehead}, {Wyrzykowski}, {Yoldas}, {{\v{Z}}erjal}, {Zucker}, {Zurbach},
  {Zwitter}, {Alecu}, {Allen}, {Allende Prieto}, {Amorim},
  {Anglada-Escud{\'e}}, {Arsenijevic}, {Azaz}, {Balm}, {Beck}, {Bernstein},
  {Bigot}, {Bijaoui}, {Blasco}, {Bonfigli}, {Bono}, {Boudreault}, {Bressan},
  {Brown}, {Brunet}, {Bunclark}, {Buonanno}, {Butkevich}, {Carret}, {Carrion},
  {Chemin}, {Ch{\'e}reau}, {Corcione}, {Darmigny}, {de Boer}, {de Teodoro}, {de
  Zeeuw}, {Delle Luche}, {Domingues}, {Dubath}, {Fodor}, {Fr{\'e}zouls},
  {Fries}, {Fustes}, {Fyfe}, {Gallardo}, {Gallegos}, {Gardiol}, {Gebran},
  {Gomboc}, {G{\'o}mez}, {Grux}, {Gueguen}, {Heyrovsky}, {Hoar}, {Iannicola},
  {Isasi Parache}, {Janotto}, {Joliet}, {Jonckheere}, {Keil}, {Kim},
  {Klagyivik}, {Klar}, {Knude}, {Kochukhov}, {Kolka}, {Kos}, {Kutka}, {Lainey},
  {LeBouquin}, {Liu}, {Loreggia}, {Makarov}, {Marseille}, {Martayan},
  {Martinez-Rubi}, {Massart}, {Meynadier}, {Mignot}, {Munari}, {Nguyen},
  {Nordlander}, {Ocvirk}, {O'Flaherty}, {Olias Sanz}, {Ortiz}, {Osorio},
  {Oszkiewicz}, {Ouzounis}, {Palmer}, {Park}, {Pasquato}, {Peltzer}, {Peralta},
  {P{\'e}turaud}, {Pieniluoma}, {Pigozzi}, {Poels}, {Prat}, {Prod'homme},
  {Raison}, {Rebordao}, {Risquez}, {Rocca-Volmerange}, {Rosen}, {Ruiz-Fuertes},
  {Russo}, {Sembay}, {Serraller Vizcaino}, {Short}, {Siebert}, {Silva},
  {Sinachopoulos}, {Slezak}, {Soffel}, {Sosnowska}, {Strai{\v{z}}ys}, {ter
  Linden}, {Terrell}, {Theil}, {Tiede}, {Troisi}, {Tsalmantza}, {Tur},
  {Vaccari}, {Vachier}, {Valles}, {Van Hamme}, {Veltz}, {Virtanen}, {Wallut},
  {Wichmann}, {Wilkinson}, {Ziaeepour}, \& {Zschocke}}]{GaiaMission_2016}
{Gaia Collaboration}, {Prusti}, T., {de Bruijne}, J.~H.~J., {et~al.} 2016,
  \aap, 595, A1, \dodoi{10.1051/0004-6361/201629272}

\bibitem[{Gaudi {et~al.}(2005)Gaudi, Seager, \& Mallen-Ornelas}]{Gaudi2005}
Gaudi, B.~S., Seager, S., \& Mallen-Ornelas, G. 2005, The Astrophysical
  Journal, 623, 472, \dodoi{10.1086/428478}

\bibitem[{Gonzalez(1997)}]{Gonzalez1997}
Gonzalez, G. 1997, Monthly Notices of the Royal Astronomical Society, 285, 403,
  \dodoi{10.1093/mnras/285.2.403}

\bibitem[{Heller(2019)}]{Heller2019}
Heller, R. 2019, Astronomy \& Astrophysics, 628, A42,
  \dodoi{10.1051/0004-6361/201833486}

\bibitem[{Hellier {et~al.}(2017)Hellier, Anderson, Cameron, Delrez, Gillon,
  Jehin, Lendl, Maxted, {Neveu-VanMalle}, Pepe, Pollacco, Queloz,
  S{\'e}gransan, Smalley, Southworth, Triaud, Udry, Wagg, \&
  West}]{Hellier2017}
Hellier, C., Anderson, D.~R., Cameron, A.~C., {et~al.} 2017, Monthly Notices of
  the Royal Astronomical Society, 465, 3693, \dodoi{10.1093/mnras/stw3005}

\bibitem[{Huang {et~al.}(2020{\natexlab{a}})Huang, Vanderburg, P{\'a}l, Sha,
  Yu, Fong, Fausnaugh, Shporer, Guerrero, Vanderspek, \&
  Ricker}]{TESS_QLP_Huang2020a}
Huang, C.~X., Vanderburg, A., P{\'a}l, A., {et~al.} 2020{\natexlab{a}},
  Research Notes of the AAS, 4, 204, \dodoi{10.3847/2515-5172/abca2e}

\bibitem[{Huang {et~al.}(2020{\natexlab{b}})Huang, Vanderburg, P{\'a}l, Sha,
  Yu, Fong, Fausnaugh, Shporer, Guerrero, Vanderspek, \&
  Ricker}]{TESS_QLP_Huang2020b}
---. 2020{\natexlab{b}}, Research Notes of the AAS, 4, 206,
  \dodoi{10.3847/2515-5172/abca2d}

\bibitem[{{Lightkurve Collaboration} {et~al.}(2018){Lightkurve Collaboration},
  {Cardoso}, {Hedges}, {Gully-Santiago}, {Saunders}, {Cody}, {Barclay}, {Hall},
  {Sagear}, {Turtelboom}, {Zhang}, {Tzanidakis}, {Mighell}, {Coughlin}, {Bell},
  {Berta-Thompson}, {Williams}, {Dotson}, \& {Barentsen}}]{Lightkurve18}
{Lightkurve Collaboration}, {Cardoso}, J.~V.~d.~M., {Hedges}, C., {et~al.}
  2018, {Lightkurve: Kepler and TESS time series analysis in Python},
  Astrophysics Source Code Library.
\newblock \doeprint{1812.013}

\bibitem[{{NASA Exoplanet Archive}(2022)}]{ExoplanetArchive_PSCompPars}
{NASA Exoplanet Archive}. 2022, Planetary Systems Composite Parameters,
  Version: 2022-02-14,  NExScI-Caltech/IPAC, \dodoi{10.26133/NEA13}

\bibitem[{Nelson {et~al.}(2017)Nelson, Ford, \& Rasio}]{Nelson2017}
Nelson, B.~E., Ford, E.~B., \& Rasio, F.~A. 2017, The Astronomical Journal,
  154, 106, \dodoi{10.3847/1538-3881/aa82b3}

\bibitem[{Petigura {et~al.}(2017)Petigura, Howard, Marcy, Johnson, Isaacson,
  Cargile, Hebb, Fulton, Weiss, Morton, Winn, Rogers, Sinukoff, Hirsch, \&
  Crossfield}]{CKSI_Petigura2017}
Petigura, E.~A., Howard, A.~W., Marcy, G.~W., {et~al.} 2017, The Astronomical
  Journal, 154, 107, \dodoi{10.3847/1538-3881/aa80de}

\bibitem[{Petigura {et~al.}(2018)Petigura, Marcy, Winn, Weiss, Fulton, Howard,
  Sinukoff, Isaacson, Morton, \& Johnson}]{Petigura2018}
Petigura, E.~A., Marcy, G.~W., Winn, J.~N., {et~al.} 2018, The Astronomical
  Journal, 155, 89, \dodoi{10.3847/1538-3881/aaa54c}

\bibitem[{{Rappaport} {et~al.}(2013){Rappaport}, {Sanchis-Ojeda}, {Rogers},
  {Levine}, \& {Winn}}]{Rappaport2013}
{Rappaport}, S., {Sanchis-Ojeda}, R., {Rogers}, L.~A., {Levine}, A., \& {Winn},
  J.~N. 2013, \apjl, 773, L15, \dodoi{10.1088/2041-8205/773/1/L15}

\bibitem[{{Recio-Blanco} {et~al.}(2022){Recio-Blanco}, de~Laverny, Palicio, \&
  Kordopatis}]{GaiaDR3_GSPSpec_Recio-Blanco2022}
{Recio-Blanco}, A., de~Laverny, P., Palicio, P.~A., \& Kordopatis, G. 2022,
  Astronomy \& Astrophysics, \dodoi{10.1051/0004-6361/202243750}

\bibitem[{{Ricker} {et~al.}(2015){Ricker}, {Winn}, {Vanderspek}, {Latham},
  {Bakos}, {Bean}, {Berta-Thompson}, {Brown}, {Buchhave}, {Butler}, {Butler},
  {Chaplin}, {Charbonneau}, {Christensen-Dalsgaard}, {Clampin}, {Deming},
  {Doty}, {De Lee}, {Dressing}, {Dunham}, {Endl}, {Fressin}, {Ge}, {Henning},
  {Holman}, {Howard}, {Ida}, {Jenkins}, {Jernigan}, {Johnson}, {Kaltenegger},
  {Kawai}, {Kjeldsen}, {Laughlin}, {Levine}, {Lin}, {Lissauer}, {MacQueen},
  {Marcy}, {McCullough}, {Morton}, {Narita}, {Paegert}, {Palle}, {Pepe},
  {Pepper}, {Quirrenbach}, {Rinehart}, {Sasselov}, {Sato}, {Seager},
  {Sozzetti}, {Stassun}, {Sullivan}, {Szentgyorgyi}, {Torres}, {Udry}, \&
  {Villasenor}}]{TESS_Ricker15}
{Ricker}, G.~R., {Winn}, J.~N., {Vanderspek}, R., {et~al.} 2015, Journal of
  Astronomical Telescopes, Instruments, and Systems, 1, 014003,
  \dodoi{10.1117/1.JATIS.1.1.014003}

\bibitem[{Rodriguez {et~al.}(2023)Rodriguez, Quinn, Vanderburg, Zhou, Eastman,
  Thygesen, Cale, Ciardi, Reed, Oelkers, Collins, Bieryla, Latham, Gaudi,
  Hellier, Sokolovsky, Schulte, Srdoc, Kielkopf, Horta, Massey, Evans,
  Stephens, McLeod, Chazov, Krushinsky, Ghachoui, Safonov, Dedrick, Conti,
  Laloum, Giacalone, Ziegler, Serra, Nogues, Murgas, Michaels, Ricker,
  Vanderspek, Winn, Jenkins, Addison, Alfaro, Anderson, Ayad, Bedding,
  Belinsky, Benkhaldoun, Berlind, Blake, Bowen, Bowler, Boyle, Branson,
  Briceno, Calkins, Campbell, Chomiuk, Collins, Cornachione, Daassou, Dressing,
  Esquerdo, Feliz, Fong, Gan, Gill, Goliguzova, Hansen, Hintz, Horner, Huang,
  James, Jensen, Johnson, Kane, Barkaoui, Kim, Kim, Kuhn, Law, Lewin, Liu,
  Lund, Mann, McCrady, Mengel, Mink, Murphy, Narita, Newman, Okumura, Osborn,
  Paegert, Palle, Pepper, Plavchan, Popov, Rabus, Ranshaw, Rodriguez, Roh,
  Reefe, Savel, Schwarz, Shporer, Siverd, Sliski, Stassun, Stevens, Soubkiou,
  Ting, Tinney, Vowell, West, Wilson, Wittenmyer, Wittrock, Wright, Zhang, \&
  Zobel}]{Rodriguez2022}
Rodriguez, J.~E., Quinn, S.~N., Vanderburg, A., {et~al.} 2023, \mnras,
  \dodoi{10.1093/mnras/stad595}

\bibitem[{Santerne {et~al.}(2016)Santerne, Moutou, Tsantaki, Bouchy,
  H{\'e}brard, Adibekyan, Almenara, Amard, Barros, Boisse, Bonomo, Bruno,
  Courcol, Deleuil, Demangeon, D{\'i}az, Guillot, Havel, Montagnier,
  Rajpurohit, Rey, \& Santos}]{Santerne2016}
Santerne, A., Moutou, C., Tsantaki, M., {et~al.} 2016, Astronomy \&
  Astrophysics, 587, A64, \dodoi{10.1051/0004-6361/201527329}

\bibitem[{Santos {et~al.}(2004)Santos, Israelian, \& Mayor}]{Santos2004}
Santos, N.~C., Israelian, G., \& Mayor, M. 2004, Astronomy \& Astrophysics,
  415, 1153, \dodoi{10.1051/0004-6361:20034469}

\bibitem[{Schlaufman(2018)}]{Schlaufman2018}
Schlaufman, K.~C. 2018, The Astrophysical Journal, 853, 37,
  \dodoi{10.3847/1538-4357/aa961c}

\bibitem[{Sha {et~al.}(2022)Sha, Vanderburg, Huang, Armstrong, Brahm,
  Giacalone, Wood, Collins, Nielsen, Hobson, Ziegler, Howell, {Torres-Miranda},
  Mann, Zhou, {Delgado-Mena}, Rojas, Abe, Trifonov, Adibekyan, Sousa,
  {Fajardo-Acosta}, Guillot, Howard, Littlefield, Hawthorn, Schmider,
  Eberhardt, Tan, Osborn, Schwarz, Str{\o}m, Jord{\'a}n, Wang, Henning, Massey,
  Law, Stockdale, Furlan, Srdoc, Wheatley, Navascu{\'e}s, Lissauer, Stassun,
  Ricker, Vanderspek, Latham, Winn, Seager, Jenkins, Barclay, Bouma,
  Christiansen, Guerrero, \& Rose}]{Sha2022}
Sha, L., Vanderburg, A.~M., Huang, C.~X., {et~al.} 2022.
\newblock \doarXiv{2209.14396}

\bibitem[{Twicken {et~al.}(2016)Twicken, Jenkins, Seader, Tenenbaum, Smith,
  Brownston, Burke, Catanzarite, Clarke, Cote, Girouard, Klaus, Li, McCauliff,
  Morris, Wohler, Campbell, Uddin, Zamudio, Sabale, Bryson, Caldwell,
  Christiansen, Coughlin, Haas, Henze, Sanderfer, \& Thompson}]{Twicken2016}
Twicken, J.~D., Jenkins, J.~M., Seader, S.~E., {et~al.} 2016, The Astronomical
  Journal, 152, 158, \dodoi{10.3847/0004-6256/152/6/158}

\bibitem[{Udry {et~al.}(2003)Udry, Mayor, \& Santos}]{Udry2003}
Udry, S., Mayor, M., \& Santos, N.~C. 2003, Astronomy \& Astrophysics, 407,
  369, \dodoi{10.1051/0004-6361:20030843}

\bibitem[{Valenti \& Fischer(2005)}]{Valenti2005}
Valenti, J.~A., \& Fischer, D.~A. 2005, The Astrophysical Journal Supplement
  Series, 159, 141, \dodoi{10.1086/430500}

\bibitem[{{Valenti} \& {Piskunov}(1996)}]{SME_Valenti1996}
{Valenti}, J.~A., \& {Piskunov}, N. 1996, \aaps, 118, 595

\bibitem[{van Cleve {et~al.}(2016)van Cleve, Howell, Smith, Clarke, Thompson,
  Bryson, Lund, Handberg, \& Chaplin}]{vanCleve2016}
van Cleve, J.~E., Howell, S.~B., Smith, J.~C., {et~al.} 2016, Publications of
  the Astronomical Society of the Pacific, 128, 075002,
  \dodoi{10.1088/1538-3873/128/965/075002}

\bibitem[{Winn {et~al.}(2017)Winn, {Sanchis-Ojeda}, Rogers, Petigura, Howard,
  Isaacson, Marcy, Schlaufman, Cargile, \& Hebb}]{Winn2017}
Winn, J.~N., {Sanchis-Ojeda}, R., Rogers, L., {et~al.} 2017, The Astronomical
  Journal, 154, 60, \dodoi{10.3847/1538-3881/aa7b7c}

\bibitem[{Yee {et~al.}(2021)Yee, Winn, \& Hartman}]{Yee2021b}
Yee, S.~W., Winn, J.~N., \& Hartman, J.~D. 2021, The Astronomical Journal, 162,
  240, \dodoi{10.3847/1538-3881/ac2958}

\bibitem[{Yee {et~al.}(2022)Yee, Winn, Hartman, Rodriguez, Zhou, Quinn, Latham,
  Bieryla, Collins, Addison, Angelo, Barkaoui, Benni, Boyle, Brahm, Butler,
  Ciardi, Collins, Conti, Crane, Dai, Dressing, Eastman, Essack,
  {For{\'e}s-Toribio}, Furlan, Gan, Giacalone, Gill, Girardin, Henning, Henze,
  Hobson, Horner, Howard, Howell, Huang, Isaacson, Jenkins, Jensen, Jord{\'a}n,
  Kane, Kielkopf, Lasota, Levine, Lubin, Mann, Massey, McLeod, Mengel,
  Mu{\~n}oz, Murgas, Palle, Plavchan, Popowicz, Radford, Ricker, Rowden,
  Safonov, Savel, Schwarz, Seager, Sefako, Shporer, Srdoc, Strakhov, Teske,
  Tinney, Tyler, Wittenmyer, Zhang, \& Ziegler}]{Yee2022a}
Yee, S.~W., Winn, J.~N., Hartman, J.~D., {et~al.} 2022, The Astronomical
  Journal, 164, 70, \dodoi{10.3847/1538-3881/ac73ff}

\bibitem[{Yee {et~al.}(2023)Yee, Winn, Hartman, Bouma, Zhou, Quinn, Latham,
  Bieryla, Rodriguez, Collins, Alfaro, Barkaoui, Beard, Belinski, Benkhaldoun,
  Benni, Bernacki, Boyle, Butler, Caldwell, Chontos, Christiansen, Ciardi,
  Collins, Conti, Crane, Daylan, Dressing, Eastman, Essack, Evans, Everett,
  {Fajardo-Acosta}, {For{\'e}s-Toribio}, Furlan, Ghachoui, Gillon, Hellier,
  Helm, Howard, Howell, Isaacson, Jehin, Jenkins, Jensen, Kielkopf, Laloum,
  {Leonhardes-Barboza}, Logsdon, Lubin, Lund, MacDougall, Mann, Maslennikova,
  Massey, McLeod, Mu{\~n}oz, Newman, Orlov, Plavchan, Popowicz, Pozuelos,
  Pritchard, Radford, Reefe, Ricker, Rudat, Safonov, Schwarz, Schweiker, Scott,
  Seager, Shectman, Stockdale, Tan, Teske, Thomas, Timmermans, Vanderspek,
  Vermilion, Watanabe, Weiss, West, Van~Zandt, Zejmo, \& Ziegler}]{Yee2022b}
---. 2023, \apjs, 265, 1, \dodoi{10.3847/1538-4365/aca286}

\bibitem[{Zhou {et~al.}(2019)Zhou, Huang, Bakos, Hartman, Latham, Quinn,
  Collins, Winn, Wong, Kov{\'a}cs, Csubry, Bhatti, Penev, Bieryla, Esquerdo,
  Berlind, Calkins, de~{Val-Borro}, Noyes, L{\'a}z{\'a}r, Papp, S{\'a}ri,
  Kov{\'a}cs, Buchhave, Szklenar, B{\'e}ky, Johnson, Cochran, Kniazev, Stassun,
  Fulton, Shporer, Espinoza, Bayliss, Everett, Howell, Hellier, Anderson,
  Cameron, West, Brown, Schanche, Barkaoui, Pozuelos, Gillon, Jehin,
  Benkhaldoun, Daassou, Ricker, Vanderspek, Seager, Jenkins, Lissauer,
  Armstrong, Collins, Gan, Hart, Horne, Kielkopf, Nielsen, Nishiumi, Narita,
  Palle, Relles, Sefako, Tan, Davies, Goeke, Guerrero, Haworth, \&
  Villanueva}]{Zhou2019a}
Zhou, G., Huang, C.~X., Bakos, G.~{\'A}., {et~al.} 2019, The Astronomical
  Journal, 158, 141, \dodoi{10.3847/1538-3881/ab36b5}

\end{thebibliography}
\bibliographystyle{aasjournal}

\end{document}